\documentclass[aps,prl,reprint,twocolumn,superscriptaddress]{revtex4-1}

\usepackage{euscript}
\usepackage{amssymb}
\usepackage{amsfonts}
\usepackage{amsbsy}
\usepackage{epsfig}
\usepackage{amsthm}
\usepackage{amscd}
\usepackage{amstext}
\usepackage{verbatim}
\usepackage{amsmath}
\usepackage{cancel}
\usepackage{capt-of}
\usepackage{empheq}
\usepackage{subfigure}
\usepackage{xcolor}

\usepackage[pdftex]{hyperref}
\hypersetup{ 
colorlinks=true, 
linkcolor=black, 
citecolor=red, 
}

\def\ben{\begin{equation}}
\def\een{\end{equation}}

\let\a=\alpha \let\b=\beta  \let\d=\delta

\let\s=\sigma

\let\pa=\partial
\def\be{\begin{equation}}
\def\ee{\end{equation}}
\def\beq{\begin{equation}}
\def\eeq{\end{equation}}
\def\ba{\begin{array}}
\def\ea{\end{array}}

\def\dalemb#1#2{{\vbox{\hrule height .#2pt
       \hbox{\vrule width.#2pt height#1pt \kern#1pt
               \vrule width.#2pt}
       \hrule height.#2pt}}}

\newcommand{\bea}{\begin{eqnarray}}
\newcommand{\eea}{\end{eqnarray}}

\newcommand{\rchi}{{\mathpalette\irchi\relax}}
\newcommand{\irchi}[2]{\raisebox{\depth}{$#1\chi$}}

\makeatletter
\newcommand*\bigcdot{\mathpalette\bigcdot@{.5}}
\newcommand*\bigcdot@[2]{\mathbin{\vcenter{\hbox{\scalebox{#2}{$\m@th#1\bullet$}}}}}
\makeatother

\renewcommand{\eqref}[1]{(\ref{#1})}

\def\R{{{\Bbb R}}}

\def\Im{{{\frak{Im}}}}

\def\ocal{{\mathcal{O}}}

\renewcommand{\Im}[0]{\operatorname{Im}}

\def \d {\partial}

\begin{document}
\frenchspacing

\title{Theory of the supercyclotron resonance and Hall response in anomalous 2d metals}

\author{Luca V. Delacr\'etaz}
\affiliation{Department of Physics, Stanford University,
Stanford, California, USA}
\affiliation{Kavli Institute for Theoretical Physics, Santa Barbara, USA}
\author{Sean A. Hartnoll}
\affiliation{Department of Physics, Stanford University,
Stanford, California, USA}
\affiliation{Kavli Institute for Theoretical Physics, Santa Barbara, USA}

\begin{abstract}

Weakly disordered superconducting films can be driven into an anomalous low temperature resistive state upon applying a magnetic field. Recent experiments on weakly disordered amorphous InO$_x$ have established that both the Hall resistivity and the frequency of a cyclotron-like resonance in the anomalous metal are highly suppressed relative to the values expected for a conventional metal. We show that both of these observations can be understood from the flux flow dynamics of vortices in a superconductor with significant vortex pinning. Results for flux flow transport are obtained using a systematic hydrodynamic expansion, controlled by the diluteness of mobile vortices at low temperatures. Hydrodynamic transport coefficients are related to microscopics through Kubo formulae for the longitudinal and Hall vortex conductivities, as well as a `vorto-electric' conductivity.

\end{abstract}

\maketitle

{\it Introduction.---} In conventional metals, the Hall resistivity and the cyclotron frequency are key observables that can often be used as proxies for the density and mass of charge carriers, respectively. Recent measurements have probed these quantities in the anomalous metallic state of weakly disordered amorphous InO$_x$ films. The low temperature superconducting state in this material becomes metallic upon applying a magnetic field greater than $H_\text{c} \approx 2 \text{T}$. In the metallic phase at a field of 5T the measured Hall resistivity is three orders of magnitude smaller than in the more conventional high temperature state \cite{Breznaye1700612}, and falls below experimental sensitivity at a lower field $H_\text{M2} > H_\text{c}$. Furthermore, the maximum of the frequency-dependent conductivity is at zero frequency to within experimental resolution \cite{PhysRevLett.111.067003}; for a conventional Drude peak this fact would require a cyclotron frequency at least four orders of magnitude smaller than expected based on normal state properties \cite{wang2017bose}. 

Anomalous metallic phases with resistive behavior similar to that of amorphous InO$_x$ have been found in many two dimensional systems, as thoroughly reviewed in \cite{review}. All of these metals emerge continuously from a superconducting phase. A rapid drop in the resistivity occurs as the temperature is lowered, before saturating to a constant at low temperatures. This suggests an interpretation of these regimes as `failed superconductors'. We will show in this letter that the experimental facts outlined above can indeed be explained by the flow of phase-disordering vortices in the would-be superconductor.

{\it Flux flow in magnetic fields.---} The extensive theoretical literature on the Hall effect due to flux flow in magnetic fields has considered a myriad of different physical effects, see e.g. \cite{PhysRev.140.A1197,doi:10.1080/14786436608211964,PhysRevLett.67.3618,PhysRevB.46.8376, Kopnin1993,  PhysRevLett.71.1242,PhysRevLett.72.3875,PhysRevLett.75.3736, PhysRevLett.83.4864}. This reflects a diverse set of experimental results in different flux flow regimes and in different materials. Much of the existing discussion involves microscopic modelling of the forces acting on vortices. We will instead argue that the diluteness of the mobile vortices allows an alternative and completely systematic treatment based on hydrodynamic argumentation combined with Kubo formulae. Our result for the Hall resistivity will be (we set $\hbar = {\text e}^\star = 1$ throughout) 
\be\label{eq:rhoxy0}
\rho_{yx} = \left(\sigma_\text{n}^H + \sigma_o^H\right) \rho_{xx}^2 + \frac{n^\text{eff}_v}{n_s} \,.
\ee
The three terms in (\ref{eq:rhoxy0}) respectively describe a Hall signal arising from currents in the vortex core, currents carried by Bogoliubov quasiparticles in the superfluid, and the co-motion of supercurrent parallel to the vortex current. The Hall conductivity $\sigma_o^H$ of the Bogoliubov quasiparticles is typically negligible due to their approximate particle-hole symmetry. Dominance of the first term, proportional to the Hall conductivity $\sigma_\text{n}^H$ of the vortex cores, leads to the relation $\rho_{yx} \sim \rho_{xx}^2$ obtained by Vinokur {\it et al.} \cite{PhysRevLett.71.1242}, and observed in some thermally activated flux flow \cite{PhysRevLett.71.617,PhysRevLett.71.3858,PhysRevLett.74.2351, PhysRevB.50.3312,PhysRevB.56.410,PhysRevB.65.184520}. This scaling arises because --- as shown by Bardeen-Stephen \cite{PhysRev.140.A1197} and rederived below --- $\rho_{xx} \sim x$, the fraction of the area occupied by mobile vortex cores, is strongly temperature and field dependent, while $\sigma_\text{n}^H$ is not. Dominance of the final `co-motion' term, on the other hand, is crucial to understand experimental results on free flux flow. There the density of vortices that co-move with the superfluid $n^\text{eff}_v \sim H$, the applied field, while $n_s$ is the superfluid density. Thus $\rho_{yx} \sim H \sim  x \sim \rho_{xx}$, as is observed \cite{PhysRevB.42.10781,PhysRevLett.73.1699}, and predicted by Nozi\`eres and Vinen \cite{doi:10.1080/14786436608211964}.

The general relationship (\ref{eq:rhoxy0}) between the Hall and longitudinal resistivities is more complicated. The first two terms in (\ref{eq:rhoxy0}) lead to $\rho_{yx} \sim \rho_{xx}^2$ while the final term leads to $\rho_{yx} \sim \rho_{xx}$, if the density of co-moving vortices $n^\text{eff}_v \sim x$. The full expression, therefore, may well explain the range of scaling relations, $\rho_{yx} \sim \rho_{xx}^\beta$ with $1 \lesssim \beta \lesssim 2$, reported in the experimental literature \cite{PhysRevLett.68.690, PhysRevLett.76.2993, CAGIGAL1995155,PhysRevB.59.R9031}. Competition between effects captured by
the first and last terms in (\ref{eq:rhoxy0}) has previously been invoked to explain the observed change in sign of the Hall response in some flux-flow regimes \cite{PhysRevB.41.11630,PhysRevB.43.6246,PhysRevB.46.8376, Kopnin1993, PhysRevLett.75.3736}.

We furthermore obtain expressions for the width $\Omega$ and frequency $\Omega_H$ of a `supercyclotron resonance' \cite{Davison:2016hno}. This resonance is due to superfluid and vortex flow in a magnetic field. It can coexist with a conventional cyclotron resonance (due to flow of the normal fluid component). We will find
\be\label{eq:Omega0}
\Omega =  \frac{2 x}{\sigma_\text{n}} f_s  \quad \text{and} \quad \Omega_H = - \frac{\pa j_v^x}{\pa u_\phi^x} \equiv \frac{n^\text{eff}_v}{m_\star} \,.
\ee
The result for $\Omega$ in (\ref{eq:Omega0}) is precisely the Bardeen-Stephen expression for vortex diffusivity \cite{PhysRev.140.A1197}, with $\sigma_\text{n}$ the conductivity of the vortex cores and $f_s$ the superfluid stiffness. In (\ref{eq:Omega0}), $\Omega_H$ is given by a static susceptibility; the 
second step in the expression defines $n^\text{eff}_v$. The mass scale $m_\star$ is such that $f_s \equiv n_s/m_\star$. Therefore the phase gradient $u_\phi \equiv \nabla \phi = m_\star v_s$, with $v_s$ the superfluid velocity. We have set $\hbar = 1$. Finally, $j_v$ is the vortex current.  With Galilean invariance, $n^\text{eff}_v = n_v$ is the full density of mobile vortices. $\Omega_H$ is then precisely the frequency appearing due to the co-motion of vortices and supercurrent in Nozi\`eres-Vinen \cite{doi:10.1080/14786436608211964}. More generally, pinning can strongly break Galilean invariance, so that the effective number of vortices that co-move with the supercurrent $n_v^\text{eff} < n_v$.

{\it Experiments on anomalous metals.---} We can return now to the measurements on InO$_x$. The first observation is that $\Omega_H \lesssim 10^{-5} \Omega$ \cite{wang2017bose}. The area occupied by mobile vortex cores is $x \sim n_v \xi^2 \sim n_v/n_s$, where $\xi$ is the superconducting correlation length. Here $x \propto n_v$ because in a magnetic field we expect the flux through all the different vortices to be aligned. Together with $\sigma_\text{n} \sim 10 \text{ e$^2$}/h \sim 0.4 \, {\text e}^{\star 2}/\hbar$ \cite{wang2017bose, Breznaye1700612}, (\ref{eq:Omega0}) then implies
\be\label{eq:nvnf}
\frac{n^\text{eff}_v}{n_v} \lesssim 10^{-5} \,.
\ee

Secondly, to a good approximation $\rho_{yx} \sim \rho_{xx}^2$ where the Hall signal is detectable \cite{Breznaye1700612}. This requires the final term in (\ref{eq:rhoxy0}) to be negligible. Thus $n^\text{eff}_v/n_v \sim n^\text{eff}_v/(n_s x) \lesssim \rho_{yx}/x \sim \rho_{yx}/(\rho_{xx} \sigma_\text{n}) = \tan \theta^H/\sigma_\text{n}$. We used the Bardeen-Stephen result $\rho_{xx} = 2x/\sigma_\text{n}$, recovered below. The measured $\tan \theta^H$ becomes as small as $10^{-4}$ \cite{Breznaye1700612}, leading to $n^\text{eff}_v/n_v \lesssim 10^{-4}$, consistent with (\ref{eq:nvnf})

The conclusion (\ref{eq:nvnf}) is therefore reached from two independent experiments. It follows that there is essentially vanishing parallel co-motion of vortices and supercurrent, as quantified by the dissipationless susceptibilty $\pa j_v^x/\pa u_\phi^x$ in (\ref{eq:Omega0}). Indeed, strong pinning in InO$_x$ causes $\rho_{xx} \sim x$ to vary by two orders of magnitude as a function of applied field in the anomalous metal \cite{Breznaye1700612}. InO$_x$ is therefore far from the `Nozi\`eres-Vinen' free flow regime.

A condition analogous to (\ref{eq:nvnf}) must also hold for the systems mentioned above where a $\rho_{yx} \sim \rho_{xx}^2$ scaling was previously observed  \cite{PhysRevLett.71.617,PhysRevLett.71.3858,PhysRevLett.74.2351, PhysRevB.50.3312,PhysRevB.56.410,PhysRevB.65.184520}. The supercyclotron resonance will be easiest to observe in materials that instead exhibit free flux flow, with negligible pinning, so that $\rho_{yx} \sim \rho_{xx}$.

The Hall resistivity measurements further reveal a weak field dependence of $\sigma_{xy}=\rho_{yx}/\rho_{xx}^2$, with $\sigma_{xy}$ possibly vanishing below a field $H_\text{M2} > H_\text{c}$ \cite{Breznaye1700612}. A strictly vanishing zero temperature $\sigma_{xy}$ over some field range requires that the vortex core contribution $\sigma^H_\text{n} = 0$ in (\ref{eq:rhoxy0}), in addition to the vanishing of vortex/superfluid co-motion implied by (\ref{eq:nvnf}). Such particle-hole symmetry \cite{Breznaye1700612} is seen away from a flux flow regime in more disordered samples \cite{Breznay12012016,PhysRevB.77.212501}.

{\it Hydrodynamic approach.---} Our analysis is anchored in the observation \cite{PhysRevLett.111.067003} of a narrow peak at zero frequency in the optical response $\sigma(\omega)$. This peak defines a lifetime that is around $10^5$ times longer than that of the electronic quasiparticles in the material. Such a hierarchy of timescales allows a systematic hydrodynamic expansion of the collective response; all non-collective modes have decayed before the timescales of interest. Furthermore, the conductance peak narrows as the magnetic field is reduced towards the onset of superconductivity at $H_\text{c}$. This strongly suggests that the appropriate low energy description of the anomalous metal is superfluid hydrodynamics with a slow phase-relaxation timescale \cite{PhysRevLett.111.067003, Davison:2016hno}. Phase relaxation requires the inclusion of vortices in the hydrodynamic description. The hydrodynamic variables are therefore the electrical and vortex currents $j$ and $j_v$ and the phase gradient $u_\phi = \nabla \phi$. The conductance peak in fact survives into the superconducting phase \cite{wang2017bose} -- at the very end we explain how this can arise from the contribution of pinned vortices to the optical conductivity.

Working within linear response and assuming homogeneous currents \footnote{Spatial inhomogeneity may be an important ingredient of anomalous metals \cite{review}. Hydrodynamics is a powerful framework for inhomogeneous dynamics, but we focus on the simplest, homogeneous situation in this work.}, the equations for the hydrodynamic variables in the presence of a uniform electric field $E$ are completely fixed. The Josephson relation, allowing for transverse vortex flow, is (with $\hbar = {\text e}^\star = 1$)
\be\label{eq:jos}
\dot u^i_\phi = E^i + \epsilon^{ij} j_v^j \,.
\ee
Here $\epsilon^{ij}$ is antisymmetric with $\epsilon^{xy}=1$. We must now express the electric and vortex currents in terms of the electric field and superfluid velocity. The most general relation that obeys the Onsager constraint 
is shown in the supplementary material to be \footnote{Here and throughout we have defined $\Omega_H$ with a different sign relative to our previous work \cite{Davison:2016hno}. We have also renamed $\rho_v$ in our previous work as $\alpha_v$, because $\rho_v$ carried misleading connotations.}:
\be
\left(
\begin{array}{c}
j_o^i \\
j_v^i
\end{array}
\right) =
\left( 
\begin{array}{cc}
\hat \sigma_o^{ij} & \hat \a^{ij}_v \\
\hat \a^{ij}_v & \hat \Omega^{ij}/f_s 
\end{array}\right)
\left(
\begin{array}{c}
E^j \\
f_s \epsilon^{jk} u^k_\phi
\end{array}
\right)
\label{eq:jv}
\ee
Here the normal component electric current $j_o \equiv j - f_s u_\phi$.
This `generalized Ohm's law' introduces six transport coefficients: $\hat \sigma_o^{ij} = \sigma_o \delta^{ij} + \sigma_o^H \epsilon^{ij}$, $\hat \Omega^{ij} = \Omega \delta^{ij} + \Omega_H \epsilon^{ij}$ and $\hat \a_v^{ij} = \a_v \delta^{ij} + \a_v^H \epsilon^{ij}$.
Below we will evaluate these coefficients through precise Kubo formulae. Their physical meaning is as follows: $\hat \Omega^{ij}$ is the vortex conductivity, $\hat \sigma_o^{ij}$ is the electrical conductivity of the normal (non-superfluid) component and $\hat \a_v$ is a `vorto-electric' conductivity.
We will drop the Hall component $\a_v^H$ in the remainder of our discussion --- its only effect on charge transport is to produce a small shift in the superfluid stiffness $f_s = n_s/m_\star$.  All of the earlier theoretical works referenced above contain equations analogous to the steady state relations (\ref{eq:jv}). We have emphasized that these equations are justified when slow phase relaxation defines a separation of timescales that allows the collective response to be treated hydrodynamically.

Solving for $j_v$ and $u_\phi$ using (\ref{eq:jos}) and (\ref{eq:jv}) gives Ohm's law $j^i = \sigma^{ij} E^j$ with 
the low-frequency conductivities \cite{Davison:2016hno}:
\begin{align}
	\sigma_{xx}(\omega) &=
		f_s \frac{(1-\a_v^2)(-i\omega + \Omega) + 2 \a_v \Omega_H}{(-i\omega + \Omega)^2 + \Omega_H^2} + \sigma_o\ , \label{eq:lonw}\\  
	\sigma_{xy}(\omega) &=
		f_s \frac{2 \a_v (-i\omega+\Omega)-(1-\a_v^2)\Omega_H}{(-i\omega + \Omega)^2 + \Omega_H^2} + \sigma_o^H\ . \label{eq:hallw}
	\end{align}
These formulae predict a `supercyclotron resonance' due to the poles at $\omega_\star = \pm \Omega_H - i \Omega$. We see that $\Omega$ determines the superfluid relaxation rate. Positivity of entropy production requires $\a_v^2 \leq \s_o \Omega/f_s$.

{\it Kubo formulae.---} The transport coefficients in (\ref{eq:lonw}) and (\ref{eq:hallw}) are given by Kubo formulae, derived in the supplementary material using the hydrodynamic Green's functions that follow from equations (\ref{eq:jos}) and (\ref{eq:jv}). The vortex conductivity is given by the retarded Green's function of the vortex current operator $J_v$. This can in turn be expressed in terms of the 
Green's function for the time derivative $\dot J_\phi = i [H, J_\phi]$ of the supercurrent:
\begin{align}
\hat \Omega^{ij} & = f_s \lim_{\omega \to 0 } \lim_{x \ll 1} \frac{1}{\omega} \Im G^R_{J^{i\vphantom{j}}_v  J^{j}_v}(\omega) \,, \label{eq:Kubo1} \\
& = f_s \left(\lim_{\omega \to 0 } \lim_{x \ll 1} \frac{1}{\omega} \Im G^R_{\dot J^{i\vphantom{j}}_\phi  \dot J^j_\phi}(\omega) - \rchi_{\dot J_\phi^{i\vphantom{j}} J^j_\phi}\right) \,.\label{eq:mem1}
\end{align}
The final contact term in (\ref{eq:mem1}) is a static susceptibility.
The `vorto-electric' conductivity depends also on the normal component current operator $J_o \equiv J - f_s J_\phi$:
\begin{align}
\a_v & = \lim_{\omega \to 0} \lim_{x \ll 1} \frac{1}{\omega} \Im G^R_{J^y_v J^y_o}(\omega) \,. \label{eq:Kubo2} \\
& = \lim_{\omega \to 0} \lim_{x \ll 1} \frac{1}{\omega} \Im G^R_{\dot J^{x \vphantom{y}}_\phi J^{y}_o}(\omega)\,. \label{eq:mem2}
\end{align}
Finally, the normal component conductivity $\hat \sigma_o^{ij}$ follows similarly from the Green's function for $J_o$. Writing the vortex conductivities in terms of $\dot J_\phi$ as in (\ref{eq:mem1}) and (\ref{eq:mem2}) will enable them to be directly related to a microscopic mechanism for phase relaxation.

In evaluating the Kubo formulae for the vortex conductivities $\hat \Omega$ and $\a_v$ it is necessary to take the limit $x \ll 1$ --- wherein mobile vortices occupy a small fraction of the sample area, ensuring slow phase relaxation --- before the zero frequency limit. This can be seen explicitly from the hydrodynamic Green's functions given in the supplementary material.
In the remainder we evaluate (\ref{eq:mem1}) and (\ref{eq:mem2}) for phase relaxation due to vortex flux flow. In \cite{Davison:2016hno} the Bardeen-Stephen phase relaxation rate $\Omega$ was recovered in this way. We can now extend that result to obtain $\Omega_H$ and $\a_v$.

{\it Supercurrent relaxation due to flux flow.---}The supercurrent operator is given by the gradient of the phase integrated {\it outside} of vortex cores, where the phase is well-defined: $J_\phi \equiv \int_{\R^2\setminus\text{cores}} \nabla \phi \, d^2x$. This definition holds in the limit of weak phase relaxation with dilute, independent vortices in an otherwise well-defined background phase -- corresponding to the $x \ll 1$ limit in the Kubo formulae, taken prior to any low frequency limit. The supercurrent operator is relaxed by charge fluctuations that are described by a `self-charging' term in the Hamiltonian: $H = \frac{1}{2\chi} \int n^2 \, d^2x$, where $n$ is the charge density and $\chi$ the charge compressibility
\footnote{This is the most relevant Hamiltonian for phase relaxation in the large core limit that we will be considering. Other terms can become relevant away from this limit.}.
The commutator $[\phi(x),n(y)] = i \delta(x-y)$ and single-valuedness of the density operator $n$ everywhere then leads to the expression
\be
\dot J_\phi = \frac{2}{\chi} \int_{\text{cores}} \nabla n \, d^2x \,. \label{eq:Jv}
\ee
This operator relation can now be used to obtain the Green's functions 
(\ref{eq:mem1}) and (\ref{eq:mem2}). The factor of 2 in (\ref{eq:Jv}) was missed in our previous work \cite{Davison:2016hno}, but is physically important. When computing $\dot J_\phi$ one must allow for the fact that the location of the core is time-dependent; in this way only mobile vortices are seen to contribute. See supplementary material for details.

The operator relation (\ref{eq:Jv}) is at the heart of our approach. Taking the expectation value of (\ref{eq:Jv}) in a state with a single large vortex and using $\langle\nabla n\rangle = \chi \nabla \mu$ in the core leads to the standard classical relation between the vortex current and the microscopic electric field $-\nabla \mu$ in the core \cite{tinkham1996introduction}.

If (i) correlations between excitations in distinct vortex cores are neglected and (ii) the vortex cores are assumed to be large compared to the mean free path of the normal state in the core, then the Kubo formulae can be evaluated explicitly. Using the operator (\ref{eq:Jv}), the first contribution to (\ref{eq:mem1}) becomes
\be
\frac{1}{\omega} \Im G^R_{\dot J^{i\vphantom{j}}_\phi  \dot J^j_\phi}(\omega) = - x \, \frac{4}{\chi^2}  \int_\text{core} \lim_{\omega \to 0} \pa_i \pa_j \frac{ \Im G^R_{n n}(\omega,y)}{\omega} d^2y \,, \label{eq:integral}
\ee
with $x$ the fraction of the total area covered by mobile vortex cores. The integral is over a single core. The control parameter in this entire computation is $x \ll 1$, so that dilute vortices lead to slow phase relaxation. The large core assumption allowed the Green's function in the core to be translationally invariant so that $G^R(x,y) = G^R(x-y)$.
In the large core limit the charge density diffuses so that $G^R_{nn}(\omega,k) = \sigma_\text{n} k^2/(- i \omega + D k^2)$. The conductivity of the normal state in the core $\sigma_\text{n} = \chi D$, with $D$ the diffusivity. The integral in (\ref{eq:integral}) is then easily evaluated to give
\be\label{eq:kubobit}
\frac{1}{\omega} \Im G^R_{\dot J^{i\vphantom{j}}_\phi  \dot J^j_\phi}(\omega) = \frac{2 x}{\sigma_\text{n}} \delta^{ij} \,.
\ee

The susceptibility term in (\ref{eq:mem1}) can be written
\be\label{eq:sus}
\rchi_{\dot J_\phi^{i\vphantom{j}} J^j_\phi} = \frac{1}{f_s} \frac{\pa \dot u_\phi^i}{\pa u_\phi^j} = \frac{\epsilon^{ik}}{f_s} \frac{\pa j_v^k}{\pa u_\phi^j} \,.
\ee
The first equality uses $\rchi_{AB} = \pa \langle A \rangle/\pa s_B$. Here $s_B$ is the source for $B$, and in the case at hand $s_{u_\phi} = f_s u_\phi$. The second equality uses the Josephson relation (\ref{eq:jos}). The electric field term, which is in fact $E - \nabla \mu$ in general, drops out because $E$ is held fixed and $\rchi_{J_\phi \nabla \mu} = 0$
at any nonzero temperature (where the response at low wavevector $k$ is nonsingular, so that $\rchi_{J_\phi \nabla_i \mu} \sim k_i \to 0$). Putting (\ref{eq:kubobit}) and (\ref{eq:sus}) together gives the results for $\Omega$ and $\Omega_H$
stated in (\ref{eq:Omega0}) above. Finally, the inclusion of correlations between distinct vortex cores and finite size corrections to Green's functions in the cores (i.e. lifting the two assumptions made above) do not lead to additional contributions to $\Omega_H$, as we note in the supplementary material.

With the same assumptions, the vorto-electric conductivity similarly gets a contribution from inside the vortex cores given by
\be\label{eq:rhov0}
\a_v =  -\frac{x}{\chi}  \int_\text{core} \lim_{\omega \to 0} \frac{1}{\omega} \Im G^R_{n \, \epsilon^{ij} \pa_i j_j}(\omega,y) d^2y \,. \ee
The contribution from outside of the cores turns out to be suppressed by powers of $x$ compared to the inside-core contribution, as we show in the supplementary material. The Green's function in the core appearing in (\ref{eq:rhov0}) again follows from the diffusive normal state dynamics. It is given by $G^R_{n \, \epsilon^{ij} \pa_i j_j}(\omega,k) = - i \omega \sigma^H_\text{n} k^2/(- i \omega + D k^2)$ and is derived in the supplementary material. Here $\sigma^H_\text{n}$ is the Hall conductivity of the normal state in the core. 
Using this Green's function we obtain
\be\label{eq:rhov}
\a_v = x \frac{\sigma_\text{n}^H}{\sigma_\text{n}} = - x \tan \theta^H_\text{n}\,.
\ee
Here $\theta^H_\text{n}$ is the Hall angle of the normal state.

{\it Conductivity and resistivity.---}Inserting the flux-flow results (\ref{eq:Omega0}) and (\ref{eq:rhov}) into the hydrodynamic expressions (\ref{eq:lonw}) and (\ref{eq:hallw}) gives the dc conductivities at small $x$:
\bea
\sigma_{xx} & = & \frac{\sigma_\text{n}}{2 x}
\,, \label{eq:lon} \\
\sigma_{xy} & = &  \sigma_\text{n}^H + \sigma_o^H + \sigma^2_{xx} \frac{n^\text{eff}_v}{n_s} \,. \label{eq:hal}
\eea
The final term in (\ref{eq:hal}) is larger than the first two by a factor of $1/x$, because $\sigma^2_{xx} \sim 1/x^2$ and $n^\text{eff}_v \sim x$. We saw in our earlier discussion, however, that the other terms can dominate when $n_v^{\rm eff}$ is suppressed.
Assuming $\sigma_{xy} \ll \sigma_{xx}$ then gives the Hall resistivity (\ref{eq:rhoxy0}).

{\it Final remark.---}The hydrodynamic theory can be extended into the superconducting phase, and explains how dynamical depinning of vortices leads to the zero frequency conductance peaks observed in \cite{wang2017bose}. Ignoring the (small) parity-odd terms, the optical conductivity (\ref{eq:lonw}) is a simple Lorentzian $\sigma(\omega) = f_s/(-i\omega + \Omega)$. We have noted that $\Omega$ is the vortex conductivity. A simple model of vortex pinning is to let $\Omega \to \Omega(\omega) = \omega \Omega/(\omega + i \omega_o)$. Here $\omega_o$ is a pinning frequency. This form arises in the limit of strong momentum relaxation from the general hydrodynamics of pinned lattices \cite{PhysRevB.96.195128}. The upshot is then the optical conductivity
\be
\sigma(\omega) = \frac{f_s}{\Omega + \omega_o} \left( \frac{\omega_o}{-i\omega} + \frac{\Omega}{- i \omega + \Omega + \omega_o} \right) \,.
\ee
A superconducting delta function arises once the pinning frequency $\omega_o$ becomes nonzero. It is accompanied by a zero frequency Lorentzian peak whose width is continuous across the superconducting-anomalous metal transition (which is driven by $\omega_o \to 0$, not $\Omega \to 0$). This is what the data shows \cite{wang2017bose}, further supporting the picture of the anomalous metal as being due to the flux flow of mobile vortices. Indeed, zero field amorphous InO$_x$ shows a canonical BKT transition as a function of temperature. The conductance peak in the high temperature BKT phase \cite{PhysRevB.84.024511} is due to mobile unpaired vortices, and is continuously connected in the phase diagram to the conductance peak seen in the anomalous metal \cite{PhysRevLett.111.067003, wang2017bose}.

\bigskip

\begin{acknowledgments}

We would especially like to acknowledge helpful input and early collaboration with Blaise Gout\'eraux. We thank Steve Kivelson and Peter Armitage for helpful comments on an earlier version of the text.
We are grateful to the hospitality of the KITP, Santa Barbara, where this work was initiated.
This research was supported in part by the National Science Foundation under Grant No. NSF PHY-1125915.

\end{acknowledgments}

%


\section*{Supplementary material}

{\it Hydrodynamics of the superfluid-vortex system.---} Hydrodynamics describes the slow excitations of a system on long lengthscales. Since total charge and vortex number are conserved, fluctuations of the densities $\delta n$ and $\delta n_v$ are slow --- and so is the dynamics of the corresponding currents $j$ and $j_v$. The observation of a long timescale associated with superconductivity \cite{PhysRevLett.111.067003} allows the phase gradient $u_\phi = \nabla \phi$ to be consistently included in the hydrodynamic description, while still  ignoring quasiparticle excitations which have a much faster characteristic timescale. We can also ignore the momentum density (and associated `normal' velocity) which decays more quickly than the phase gradient due to broken translation invariance -- this is the experimental fact that the low frequency conductance peak is purely due to phase-relaxed superconductivity \cite{PhysRevLett.111.067003}. Finally, we ignore thermal fluctuations which do not influence the charge transport results and can be straightforwardly incorporated in order to compute thermoelectric transport \cite{Davison:2016hno}.

We exclusively consider spatially homogeneous flows. The density fluctuations $\{\delta n,\delta n_v\}$ therefore drop out of the linear-response hydrodynamic equations, because they always appear differentiated. The remaining hydrodynamic variables are $\{j, j_v, u_\phi\}$.
The equilibrium Josephson relation
\begin{equation}\label{eq_Joseph}
\dot u^i_\phi = 
\nabla^i \dot \phi = E^i + \epsilon^{ij} j^j_v\, \,,
\end{equation}
is exact for homogeneous flows. $E$ is a non-dynamical background electric field. Similarly, the most general constitutive relations for the homogeneous currents, within linear response, are
\begin{align}
j^i - f_s u_\phi^i
	&=  \hat \b^{ij}_v \epsilon^{jk} (f_s u_\phi^k+ s_\phi^k) + \hat \sigma_o^{ij} E^j + \a^{ij}_v E_v^j\, , \label{eq_consti} \\
j_v^i
	&= \frac{1}{f_s}\hat \Omega^{ij}\epsilon^{jk}  (f_s u_\phi^k+ s_\phi^k) + \hat \a^{ij}_v E^j + (\hat\sigma_o^v)^{ij} E_v^j \nonumber \, . 
\end{align}
The hatted matrices have the form $\hat M^{ij} = M\delta^{ij} + M_H \epsilon^{ij}$, due to isotropy. We have introduced background sources $s_\phi,\, E=\d_t A,\, E_v=\d_t A_v$ that couple to the hydrodynamic variables as
\begin{equation}
\delta H = -\int d^2 x \, \left( s_\phi \cdot u_\phi + A\cdot j + A_v \cdot j_v \right) \, .
\end{equation}
These sources will shortly allow us to obtain the retarded Green's functions by simply studying linear response.

Although both inversion and time-reversal symmetries are broken by the magnetic field, their combination is a symmetry of the system. This leads to Onsager relations on the retarded Green's functions
\begin{equation}
G^R_{AB}(\omega) = \eta_A \eta_B G^R_{BA}(\omega)\, ,
\end{equation}
where $\eta_A$ is the sign of operator $A$ under this symmetry. For instance $\eta_{n}=\eta_{n_v}=+1$ and $\eta_\phi=-1$. The Onsager relations are then seen to impose the following relations on \eqref{eq_consti}:
\begin{equation}\label{eq:iden}
\hat\b_v = \hat \a_v \qquad \hbox{and} \qquad
\frac{1}{f_s}\hat \Omega = \hat \sigma_o^v\, .
\end{equation}
This last equation loosely identifies the phase relaxation rate as a vortex conductivity; this relation will be made more precise with a Kubo formula below. The equations (\ref{eq_consti}) together with the identifications in  (\ref{eq:iden}) are precisely (\ref{eq:jv}) in the main text, where we set the additional sources $E_v = s_\phi = 0$. As explained in the main text, we put $\a_v^H = 0$ in the remainder.

Solving the equations above for the hydrodynamic variables $\{u_\phi,\,j,\,j_v\}$ in terms of the sources $\{s_\phi,\,A,\,A_v\}$ then gives the hydrodynamic Green's functions, e.g.
\begin{equation}\label{eq_hydro_GF}
G^R_{J_\phi^i J_\phi^j}(\omega) = \frac{\d u_\phi^i(\omega)}{\d s_\phi^j(\omega)} \, .
\end{equation}
This leads to
\begin{subequations}\label{eq:appG}
\begin{align}\label{eq_GR_jvjv}
G^R_{J^{i\vphantom{j}}_v J^{j}_v}(\omega)
	&= \omega^2 G^R_{J_\phi^i J_\phi^j}(\omega) = \frac{\omega^2}{f_s} \left[\frac{\hat\Omega}{\hat\Omega-i\omega}\right]_{ij}\, ,\\
G^R_{J_v^{i\vphantom{j}} J_o^j}(\omega) \label{eq_GR_jvj0}
	&= i\omega  \epsilon^{ik} \left[G^R_{J_\phi^{k\vphantom{j}} J_o^j}(\omega)+\delta^{kj}\right]
	=  \a_v \left[\frac{\omega^2}{\hat\Omega -i\omega}\right]_{ij} \, ,
\end{align}
\end{subequations}
where we used matrix notation to make the expressions concise, for example
\begin{equation}
\left[\frac{1}{\hat\Omega -i\omega}\right]_{ij}
	= \frac{(\Omega-i\omega)\delta_{ij} + \Omega_H \epsilon_{ij}}{(\Omega-i\omega)^2 + \Omega_H^2} \, .
\end{equation}
As in the main text, we use $J_o, J_v$ and $J_\phi$ to denote the operators whose expectation values are the hydrodynamic variables $j-f_s u_\phi, j_v$ and $u_\phi$, respectively.

Using the Green's functions identities, familiar from the `memory function' description of transport \cite{PhysRevB.6.1226},
\begin{subequations}\label{eq_GFid}
\begin{align}
G^R_{\dot AB}(\omega) 
	&= -i\omega G^R_{AB}(\omega) + \chi_{\dot AB} \,, \\
G^R_{\dot A \dot B}(\omega)
	&= \omega^2 G^R_{AB}(\omega) + i\omega \chi_{\dot AB} + \chi_{\dot A\dot B}\, ,
\end{align}
\end{subequations}
where the susceptibility $\chi_{AB}\equiv G^R_{AB}(i0^+)$, eq.~\eqref{eq_GR_jvjv} leads to a Kubo formula for the phase relaxation rates
\begin{equation}\label{eq:AA}
\begin{split}
\frac{1}{f_s} \Omega_{ij}
	&= \lim_{\omega\to 0}\lim_{x \ll 1} \frac{1}{\omega}\Im G^R_{J^{i\vphantom{j}}_v J^j_v}(\omega) \\
	&= \lim_{\omega\to 0}\lim_{x \ll 1} \frac{1}{\omega}\Im G^R_{\dot J^{i\vphantom{j}}_\phi  \dot J^j_\phi}(\omega) - \chi_{\dot J_\phi^{i\vphantom{j}} J^j_\phi}\, .
\end{split}
\end{equation}
Here the order of limits just means that the components of $\hat \Omega$ in the denominators of (\ref{eq:appG}) are sent to zero before the frequency $\omega$. Taking the opposite order of limits gives zero. Similarly, eq.~\eqref{eq_GR_jvj0} yields a Kubo formula for the vorto-electric conductivity
\begin{equation}\label{eq:BB}
\begin{split}
\a_v \delta_{ij}
	&= \lim_{\omega\to 0}\lim_{x \ll 1} \frac{1}{\omega}\Im G^R_{J_v^{i\vphantom{j}} J_o^j}(\omega) \\
	&= -\epsilon^{ik} \lim_{\omega\to 0}\lim_{x \ll 1} \left[\frac{1}{\omega}\Im G^R_{\dot J^{k\vphantom{j}}_\phi J_o^j}(\omega) - \delta^{kj}\right]\, .
\end{split}
\end{equation}

Equation (\ref{eq:AA}) gives equations (\ref{eq:Kubo1}) and (\ref{eq:mem1}) in the main text. Equation (\ref{eq:BB}) gives equations (\ref{eq:Kubo2}) and (\ref{eq:mem2}) in the main text (upon putting $i=j=y$ and using isotropy).

{\it  Flux flow and supercurrent relaxation.---} We have seen that the operator $\dot J_\phi$ controls the transport coefficients $\Omega,\,\Omega_H$ and $\a_v$. Here we determine this operator in the presence of flux flow.

The superfluid phase $\phi$ is not well defined in the vortex cores. The total supercurrent operator is therefore
\begin{equation}
J_\phi = \int_{\bar c(t)} \nabla\phi\, \, ,
\end{equation}
where the integral runs over the area $\bar c(t)$ outside of the vortex cores (so that we can write $\bar c(t)\cup c(t) = \mathbb R^2$). It is important to allow these areas to depend on time, given that the phase-relaxing vortices will be mobile. The results do not depend sensitively on the assumption of a sharp vortex core boundary.

The leading effective long-wavelength Hamiltonian for charge fluctuations is $H = \frac{1}{2\chi} \int (\delta n)^2 d^2 x$, where $\chi$ is the charge susceptibility. Other local Hamiltonians for $\delta n$ lead to qualitatively similar contributions \cite{Davison:2016hno}. Therefore, using $[\phi(x),n(y)] = i \delta(x-y)$,
\begin{equation}
\begin{split}
\dot J^i_\phi 
	&= \d_t J^i_\phi + i[H,J^i_\phi]\\
	&= -\oint_{\d c(t)} dz \, v_L\cdot \hat n\,  \nabla^i \phi
	- \frac{1}{\chi}\int_{\bar c(t)}\nabla^i n \, d^2 x\\
	&\simeq \frac{1}{2} \epsilon^{ij} J^j_v- \frac{1}{\chi}\int_{\bar c(t)}\nabla^i n \, d^2 x \\ 
	&= \frac{1}{2} \dot J^i_\phi - \frac{1}{\chi}\int_{\bar c(t)}\nabla^i n \, d^2 x \, ,
\end{split}
\end{equation}
where $v_L$ is the velocity of the vortex core and $\hat n$ is a unit normal. In the third line, we set $\nabla\phi$ to its background profile around the vortex (i.e. $\phi = N \theta$ for a vortex of winding $N$) and ignored its fluctuations, since $v_L$ is already linear in fluctuations ($v_L$ has vanishing expectation value in equilibrium). As throughout, the vortex current $J_v$ is normalized so that a flux of vortex number $N$ induces a rate of change $N$ in the transverse supercurrent, rather than $2 \pi N$. In the final line we used the fact, from e.g. \eqref{eq_Joseph} or (\ref{eq_GR_jvjv}),
that $\dot J^i_\phi$ and $\epsilon^{ij} J^j_v$ are the same, up to contact terms.
Writing the last term as an integral over the cores (because the integral of $\nabla n$ over all space vanishes), we obtain the following operator equation
\begin{equation}\label{eq_Jphi_dot}
\dot J_\phi
	=  \frac{2}{\chi} \int_{c(t)} \nabla n \, d^2 x \, .
\end{equation}
This is equation (\ref{eq:Jv}) in the main text. One can easily check that the expectation values of both sides match in a classical vortex configuration, c.f. \cite{tinkham1996introduction}. Note that only mobile vortices, with a nonzero $v_L$, contribute to $\dot J_\phi$.

{\it  Evaluation of Kubo formulae.---} Setting aside the contact term in (\ref{eq:mem1}) for the moment, the Kubo formula \eqref{eq:Kubo1} can now be evaluated using \eqref{eq_Jphi_dot}
\begin{equation}\label{eq:hatO}
\hat\Omega_{ij}=\frac{4 f_s}{\chi^2V}  \int_c d^2 x \int_c d^2 y \lim_{\omega\to 0} \d_{x^i} \d_{y^j} \frac{\Im G^R_{nn}(x,y;\omega)}{\omega}\, .
\end{equation}
Here $V$ is the total volume. The contribution when $\vec x$ and $\vec y$ are in the same core is
\begin{equation}\label{eq_Omega}
\hat\Omega_{ij}= n_\text{mob} \, \frac{4f_s}{\chi^2} \int_{c_1}\! d^2 x \int_{c_1}\! d^2 y \lim_{\omega\to 0} \d_{x^i} \d_{y^j} \frac{\Im G^R_{nn}(x,y;\omega)}{\omega}\, ,
\end{equation}
where $c_1$ denotes a single vortex core and $n_\text{mob}$ is the density of mobile vortices (counted without signs; in a magnetic field we expect $n_\text{mob} = n_v$). If the cores are sufficiently large such that $G^R_{nn}(x,y;\omega) = G^R_{nn}(x-y;\omega)$, then \eqref{eq_Omega} is symmetric under $i \leftrightarrow j$ and we immediately obtain $\Omega_H = 0$. In fact, corrections due to a finite core size do not lead to a nonzero $\Omega_H$: the charge density Green's function contains no parity odd transport coefficient as long as the charge density satisfies an isotropic partial differential equation. (The current density Green's function does contain such a coefficient.)

Translation invariance of large cores means that we can work with the Fourier transform $G^R_{nn}(\omega,k)$ in the cores. This has a standard diffusive form if the vortex cores are large compared to the mean free path of the normal state in the core. This form is quoted in the main text and in (\ref{eq:nnnn}) below. Equation \eqref{eq_Omega} then becomes
\begin{equation}
\hat\Omega_{ij}
	=f_s\frac{2 x}{\sigma_{\rm n}} \delta_{ij}\, ,
\end{equation}
where $x = n_\text{mob} A_v$ is the fraction of whole area occupied by mobile vortices ($A_v$ is the surface area of a vortex core). This is equation (\ref{eq:Omega0}) in the main text, which also includes contribution to $\Omega_H$ from the contact term (\ref{eq:mem1}).

Correlations between $\vec x$ and $\vec y$ in distinct cores in (\ref{eq:hatO}) do not contribute to $\Omega_H$ -- at least to order $x^2$, where such effects would most obviously compete with the other contributions to the Hall resistivity (\ref{eq:rhoxy0}). The leading contribution (which is $\ocal(x^2)$ due to separate integrals over the areas of the two cores) vanishes, again because the charge density Green's functions do not contain any parity odd coefficients. Away from the large core limit, there can be inter-core contributions at order $x^2$ due to interactions mediated by operators other than the charge density.

One can similarly evaluate the Kubo formula \eqref{eq:Kubo2} using \eqref{eq_Jphi_dot} to find
\begin{equation}\label{eq_rhov_ff}
\a_v
 =\lim_{\omega\to 0} \frac{n_\text{mob}}{\chi} \int_{\rm c_1} d^2 x \int {d^2 y} \, \epsilon^{ij} \d_{x^i}  \frac{\Im G^R_{n\,j^0_j}(x,y;\omega)}{\omega}\, ,
\end{equation}
which receives two contributions $\a_v=\a_v|_{cc}+\a_v|_{c\bar c}$, according to whether the point $y$ is in the core $c_1$ or outside the core respectively. When $y$ is in the core $c_1$, one can again use the normal state Green's function in the core, given in the main text and in (\ref{eq:Jn}) below. Using the same approximations as above, this leads to
\begin{equation}
\a_v|_{cc} 
    = -\frac{x}{\chi} \int_{c_1} \lim_{\omega\to 0} \frac{1}{\omega} \Im G^R_{n\nabla\times j}(\omega,y) d^2y
    =x \frac{\sigma_{\rm n}^H}{\sigma_{\rm n}}\, .
\end{equation}
The situation is more complicated when $y$ is outside the core, since the Green's function solves different partial differential equations in the core near $x$ and in the superconductor near $y$. The correct procedure is to first use the core Green's function to find the charge density $\langle\delta n(x',\omega)\rangle$ on the boundary $x'\in \d c_1$ sourced by a chemical potential $\delta \mu$ localized in the core. The partial differential equations outside the cores (those of unrelaxed superfluid hydrodynamics) are then solved with boundary conditions assuming continuity of the charge density, and its derivative, at the core boundary. One can show that this procedure leads to the following prescription
\begin{equation}\label{eq_prodG}
    G^R_{nj^0}(x,y;\omega)
    \simeq
    G^{R,s}_{nj^0}(x,y;\omega)\frac{G^{R,c}_{nn}(x',y;\omega)}{G^{R,s}_{nn}(x',y;\omega)}\, ,
\end{equation}
where $G^{R,c}$ and $G^{R,s}$ denote Green's functions in the cores and superfluid respectively, and $x'$ is any point on the boundary of the vortex. This equation is only strictly valid when $y$ is at the center of the vortex core and $x$ is far away from from the core, but corrections to this equation will vanish when taking the $\omega,k\to 0$ limits associated with large cores and dilute vortices. Using the expression \eqref{eq_prodG} in \eqref{eq_rhov_ff} one finds finally
\begin{equation}
\a_v|_{c\bar c} \sim {x}^{3/2}\frac{\sigma_o^H}{\sigma_n} \,,
\end{equation}
where we dropped some geometrical factors. Thus to leading order in $x$ we find
\begin{equation}
\a_v =x \frac{\sigma_{\rm n}^H}{\sigma_{\rm n}} + \ocal({x}^{3/2}) \,,
\end{equation}
which is (\ref{eq:rhov}) in the main text.

{\it Green's functions in the vortex core.---} The vortex cores are in the normal state, and therefore have no superfluid component $f_s=0$. The constitutive relation for the current is therefore simply
\begin{equation}\label{eq_consti_inco}
j^i = \hat\sigma_{\rm n}^{ij} (E^j - \frac{1}{\chi}\nabla^j n) + \cdots \, , 
\end{equation}
where we kept one more term in the gradient expansion compared to Eq.~\eqref{eq_consti}, and $\cdots$ denotes higher order terms in this expansion. The reason we keep the leading gradient correction is that the core has finite size, and we are therefore evaluating Green's functions at finite wavevector $k$. The current Green's function can be computed by using Eq.~\eqref{eq_consti_inco} and the continuity relation $\dot n + \nabla \cdot j = 0$ to solve for $n$ and then employing \eqref{eq_hydro_GF}. It is given by
\begin{equation}
G^R_{J_i J_j} (\omega,k)
	= i\omega \left[\hat \sigma_{\rm n}^{ij} - \frac{1}{\chi}\frac{(\hat \sigma_{\rm n}\cdot k)_i (k\cdot \hat \sigma_{\rm n})_j}{-i\omega  + D k^2}\right]
\end{equation}
where the diffusion constant is $D=\sigma_{\rm n}/\chi$. The remaining Green's functions can be obtained from the Ward identity
\begin{subequations}
\begin{align}\label{eq:Jn}
G^R_{n J_j}(\omega,k)
	&= \frac{k^i}{\omega} G^R_{J_iJ_j}(\omega,k) \, , \\
G^R_{n n}(\omega,k)
	&= \frac{k^ik^j}{\omega^2} G^R_{J_iJ_j}(\omega,k) \, . \label{eq:nnnn}
\end{align}
\end{subequations}

\end{document}